# pVACview: an interactive visualization tool for efficient neoantigen prioritization and selection

Huiming Xia[1,2], My Hoang[1], Evelyn Schmidt[1], Susanna Kiwala[1], Joshua McMichael[2], Zachary L. Skidmore[1,2], Bryan Fisk[1,2], Jonathan J. Song[1,2], Jasreet Hundal[2], Thomas Mooney[2], Jason R. Walker[2], S. Peter Goedegebuure[3,4], Christopher A. Miller[1,4], William E. Gillanders[3,4], Obi L. Griffith[1,2,4,5,+], Malachi Griffith[1,2,4,5+]

Affiliations:
1. Division of Oncology, Department of Medicine, Washington University School of Medicine, St. Louis, MO, USA
2. McDonnell Genome Institute, Washington University School of Medicine, St. Louis, MO, USA
3. Department of Surgery, Washington University School of Medicine, St. Louis, MO, USA
4. Siteman Cancer Center, Washington University School of Medicine, St. Louis, MO, USA
5. Department of Genetics, Washington University School of Medicine, St. Louis, MO, USA

[+] Corresponding authors

## Abstract (350 words)

Background:
Neoantigen targeting therapies including personalized vaccines have shown promise in the treatment of cancers, particularly when used in combination with checkpoint blockade therapy. At least 100 clinical trials involving these therapies are underway globally. Accurate identification and prioritization of neoantigens is highly relevant to designing these trials, predicting treatment response, and understanding mechanisms of resistance. With the advent of massively parallel DNA and RNA sequencing technologies, it is now possible to computationally predict neoantigens based on patient-specific variant information. However, numerous factors must be considered when prioritizing neoantigens for use in personalized therapies. Complexities such as alternative transcript annotations, various binding, presentation and immunogenicity prediction algorithms, and variable peptide lengths/registers all potentially impact the neoantigen selection process. There has been a rapid development of computational tools that attempt to account for these complexities. While these tools generate numerous algorithmic predictions for neoantigen characterization, results from these pipelines are difficult to navigate and require extensive knowledge of the underlying tools for accurate interpretation. This often leads to over-simplification of pipeline outputs to make them tractable, for example limiting prediction to a single RNA isoform or only summarizing the top ranked of many possible peptide candidates. In addition to variant detection, gene expression and predicted peptide binding affinities, recent studies have also demonstrated the importance of mutation location, allele-specific anchor locations, and variation of T-cell response to long versus short peptides. Due to the intricate nature and number of salient neoantigen features, presenting all relevant information to facilitate candidate selection for downstream applications is a difficult challenge that current tools fail to address.

Results:
We have created pVACview, the first interactive tool designed to aid in the prioritization and selection of neoantigen candidates for personalized neoantigen therapies including cancer vaccines. pVACview has a user-friendly and intuitive interface where users can upload, explore, select and export their neoantigen candidates. The tool allows users to visualize candidates across three different levels, including variant, transcript and peptide information.

Conclusions:
pVACview will allow researchers to analyze and prioritize neoantigen candidates with greater efficiency and accuracy in basic and translational settings The application is available as part of the pVACtools pipeline at [pvactools.org](pvactools.org) and as an online server at [pvacview.org](pvacview.org).

## Background

Neoantigens are unique peptide sequences generated from somatic variants in tumors. These antigens provide an avenue for tumor-specific immune cell recognition and have been found to be important targets for cancer immunotherapies [1–3]. Effective neoantigens, presented by the major histocompatibility complex (MHC) and thus introduced to the patient's immune system, can prime and activate CD8+ and CD4+ T cells for downstream signaling of cell-death. Previous studies have shown that patients with relatively high tumor mutation burden tend to have stronger responses to neoantigen based immunotherapy treatments [4–6]. With the advent of massively parallel DNA and RNA sequencing technologies, researchers and clinicians are now able to computationally predict neoantigens for experimental studies of T cell biology in cancer or for design of personalized neoantigen therapies based on patient-specific mutations. Examples of such therapies include personalized neoantigen vaccines [2,7], TCR mimic antibodies [8,9], personalized adoptive T cell therapies [10,11] and engineered T cell therapies [12–14]. The process of designing bespoke neoantigen targeting therapies entails: sequencing (WGS/WES, RNAseq) of matched tumor-normal samples, somatic variant calling together with germline variant calling and HLA typing, neoantigen prediction, and selection of neoantigen candidates for manufacturing (**Figure 1**).

Numerous aspects of the process of neoantigen generation and presentation must be considered for effective target selection (**Figure 2**). These aspects include but are not limited to: 1) neoantigen mutation identification and expression, 2) peptide processing and transport, 3) peptide-MHC binding, 4) peptide-MHC stability, 5) recognition by cytotoxic T cells [15]. Additional considerations relate to manufacturability, which vary by therapeutic platform, and safety considerations including stability of the formulation and potential for off-target effects. There has been a rapid development of computational tools in an attempt to account for these complexities (**Additional file 1**). Pipelines have been developed to allow researchers to run an ensemble of many tools for individual patients, generating more than 118 features, which include metrics such as algorithmic predictions of binding, allele frequency and expression, similarity to a reference proteome, and others [16–19]. However, the results from these complex pipelines are often overwhelming in number, difficult to navigate, and require extensive knowledge of the underlying tools for accurate interpretation. Though gene expression and predicted peptide binding affinities are common features of most approaches, recent studies have also shown the importance of mutation location, allele-specific anchor locations, the potential impact of multiple class I/II short peptides arising from a single mutation, and the variation of T-cell response to long versus short peptides [20–23]. These additional complexities can be difficult to interrogate directly from computational pipeline outputs, if they are available at all.

With the high cost of personalized neoantigen therapies and difficulties in accurate neoantigen prioritization, it is critical to provide multiple levels of information to support the most effective and efficient neoantigen target selection. As with most personalized treatments, choosing the right approach depends on integrating clinical information and observations with genomic data and algorithmic outputs. Supporting this effectively means moving beyond static reports and building dynamic interfaces that provide layered information in an intuitive manner. To address these concerns, we built a comprehensive visualization tool, pVACview, that takes complex neoantigen candidate information as input, visualizes the output with multiple levels of detail, and exports results and annotations for further review and manufacturing for diverse downstream applications, including personalized neoantigen vaccines. Our previously published cancer immunotherapy pipeline, pVACtools [16], generates numerous features for neoantigen characterization. However, these outputs also require extensive additional analysis due to the intricate nature of neoantigen features. With pVACview, we now provide a complete neoantigen detection and design pipeline. The application is compatible with human, mouse and canine data and has been used in the setting of several cancer immunotherapy clinical trials (e.g. NCT04397003, NCT03422094, NCT04015700 etc.). It also supports visualization of candidate neoantigens from alternative pipelines such as vaxrank [24] and annotation tools such as NeoFox[17].

## Implementation

pVACview is written in R and is implemented as part of pVACtools, which is a computational toolkit that helps identify and visualize neoantigen candidates [16,25]. While pVACview can be used as a stand-alone tool (see Overall architecture of the software implementation), we recommend using pVACtools to generate the required inputs to access the maximum functionality. Code changes are integrated using GitHub pull requests (https://github.com/griffithlab/pVACtools/pulls). Documentation is hosted on Read the Docs (readthedocs.org) and can be viewed at https://pvactools.readthedocs.io/en/latest/pvacview.html.

A demonstration data set is provided and consists of Class I and Class II neoantigen candidate files generated from the HCC1395 breast cancer cell line and its matched lymphoblastoid cell line HCC1395BL (please refer to data availability section). The tumor and normal datasets were processed using an immunogenomics pipeline written in WDL (immuno.wdl available at https://github.com/wustl-oncology/analysis-wdls). This pipeline accepts raw tumor/normal exome and tumor RNA-seq data in FASTQ or unaligned BAM format and performs alignment, HLA typing, germline variant calling, somatic variant calling, variant phasing, variant annotation, expression analysis, RNA fusion detection and neoantigen identification. Aggregated neoantigen reports and metrics files used as inputs to pVACview were generated using the `pvacseq generate_aggregate_report` command. These datasets are available at https://github.com/griffithlab/pVACtools/tree/latestpvactools/tools/pvacview/data.

To acquire pVACtools output (specifically, pVACseq output) for use with pVACview, users can run pVACseq from the command line using variants from their own pipeline (in VCF format), or start with raw sequence data and use an end-to-end pipeline on the cloud by launching our pre-configured workflow on Dockstore (https://dockstore.org/workflows/github.com/griffithlab/analysis-wdls/immuno) via various platforms (DNAnexus, Terra, eLazi, AnVIL, NHLBI BioData Catalyst). A guide for employing the pre-configured immuno workflow to run pVACtools on Terra as an example is available at https://workflow-course.pvactools.org/index.html.

**Overall architecture of the software implementation**

pVACview has 3 modules: 1) pVACview main module, 2) NeoFox and 3) custom module. Comparison of level of details for some of these inputs can be reviewed in **Additional file 2**.

**pVACview main module**
The pVACview main module is split into the following components: user data upload, neoantigen feature visualization and exploration, and export of prioritized neoantigens and associated annotations for downstream applications **(Figure 3)**. Below, we step through each individual section in detail.

**Configuration and data import**
Generation of the neoantigen candidate input files requires preprocessing using pVACseq starting from patient samples' variant information (supplied as a VCF file). pVACseq produces neoantigen candidates with numerous features to be considered during prioritization. Two of pVACseq's output files, an aggregated candidate file (tsv format) and a metrics file (json format), serve as input files to pVACview. The aggregated candidate file contains a list of all variants with summary level information such as: best predicted neoantigen candidate and its prediction score, DNA/RNA depth and variant allele frequencies, gene and allele expression, etc. (for further details please refer to the online documentation at pvactools.org). The metrics json file contains extensive additional transcript and peptide-level information that is needed for certain features of the pVACview application.

Users have the option to additionally include a tsv file with supplemental candidate information from a different set or class of HLA alleles. This allows users to view basic median binding information of Class II results while looking at detailed Class I prediction results or vice versa. For users investigating a specific gene set of their own interest, we provide the option of uploading a tsv file where each line contains an individual gene name (e.g. names of known cancer driver genes). These genes, if found in the aggregate report file, will be highlighted in a green box with bold font in the Gene report column of the visualization interface.

**Neoantigen visualization and exploration**
Uploaded neoantigen candidates can be explored and analyzed in several different ways. Users are provided with neoantigen features that are organized into three levels of detail: variant-level, transcript-level, and peptide-level.

Variant-level information is presented in the main aggregate report table, showcasing the best neoantigen candidate for each variant as well as genomic information (e.g. gene identifier, amino acid change and position of the variant within the core binding peptide), expression level, DNA and RNA variant allele frequency, median binding prediction scores, and the total number of peptides beyond the best one that meet specified cutoffs. Each variant in the main aggregate report table is assigned a Tier based on suitability for treatment manufacture on criteria including binding affinity, expression, transcription support level, clonality, and anchor scenario. By default, the variants in this table are ordered based on their assigned tier.

Once a specific variant is selected, users are provided with a variant and gene info box, which provides further information on the exact genomic location and nucleic acid change. We have also included a link to the OpenCRAVAT variant report for the respective variant[26]. This report allows users to explore rich variant information including: variant effect annotations, associated cancer types, population allele frequencies, clinical relevance, gene annotation, pathogenicity predictions, etc.

Additionally, users are provided with individual transcripts containing the variant. The selected variant may occur within multiple transcripts that may result in distinct neoantigen peptide sequences and those that produce good binding predictions against at least one HLA allele are shown in the transcript table. The expression level of each transcript is provided as further guidance when selecting the best neoantigen candidate. In some cases, transcript sequence context impacts the peptide sequence surrounding a variant (e.g. nearby exon-intron boundaries as depicted in Figure 2). Multiple transcripts that give rise to the exact same list of peptide candidates are grouped into a single transcript set and those that give rise to different peptides are grouped into distinct transcript sets. By selecting a specific transcript set, users are then provided with a peptide table. The peptide table displays all peptide sequences from your selected transcript that were predicted to be good binders (for at least one HLA allele). Both mutant (MT) and wildtype (WT) sequences are shown, along with median binding affinities (where the MT score passed the binding threshold).

Users are also provided with binding and anchor information for each individual peptide. By selecting each pair of MT/WT peptides, users can access: 1) plots of the individual IC50-based binding affinity predictions of the MT and WT peptides for HLA alleles with at least one predicted strong binder 2) plots of the individual percentile-based binding affinity predictions 3) a binding affinity table with numerical IC50 and percentile values for each algorithm and 4) a table of orthogonal predictions trained on peptide elution mass spectrometry (e.g. BigMHC_EL, MHCFlurryEL, NetMHCPanEL) and immunogenicity data (e.g. BigMHC_IM, DeepImmuno). Note that each peptide may have up to 8 binding algorithm scores for Class I alleles (with pVACseq version 3.0 or higher) or up to 4 binding algorithm scores for Class II alleles. These views facilitate evaluation of algorithm concordance and integration of predictions pertaining to MHC binding, processing and immunogenicity.

For each peptide, we also provide users with an allele-specific anchor prediction heatmap, based on computational predictions from our previous work[23] . These predictions are normalized probabilities representing the likelihood of each position of the peptide to participate in anchoring to the HLA allele. The top 30 MT/WT peptide pairs from the peptide table are shown with anchor probabilities overlaid as a heatmap. The anchor probabilities shown are both allele and peptide length specific. In the anchor heatmap view, the mutated amino acid is marked in red (for missense mutations) and each MT/WT pair is separated from others using a dotted line. Finally, the thresholds used for determining allele specific anchors sites are provided as an anchor weights table.

After consulting the breadth of information displayed in pVACview, users can label each variant with their corresponding evaluation using the dropdown menu in the "Eval" column of the aggregate report view. The number of evaluations performed (accept, reject, review) are tracked in the peptide evaluation overview section beneath the aggregate report view.

**Export of neoantigen evaluations and final report**
When users have either finished ranking neoantigen candidates or need to pause and would like to save current evaluations, they can export the current main aggregate report using the export page. We provide two download file types (tsv and excel). The excel format is user-friendly for downstream visualization and manipulation. However, if the user plans to continue editing the aggregate report and would like to load it back in pVACview with the previous evaluations preloaded, they must use the tsv format. This serves as a way to save progress as all evaluations are cleared upon closing or refreshing the pVACview app.

**NeoFox module**

**Data import**
pVACview also takes the output of the neoantigen annotation pipeline NeoFox [17], as input. NeoFox output is a tab-separated file, where each row is one neoantigen candidate. The NeoFox format also optionally supports annotation of each candidate with a  Patient identifier and gene-level information (gene name, DNA/RNA allele frequencies). The peptide level information generated by NeoFox is comprehensive including mutated and wild type peptide sequence, mutated position, and scores for ranking peptides based on 17 neoantigen features and prediction algorithms including several that are not otherwise supported by pVACtools directly including: recognition potential, generator rate, PRIME and HEX.

**Neoantigen visualization and exploration**
pVACview provides three tabs for NeoFox data exploration. The first tab, 'Annotated Neoantigen Candidates using NeoFox' will show all neoantigen candidates and their corresponding information from the input. In the second tab 'Data Visualization', users can select up to 6 information categories to visualize the neoantigens in the form of violin plots. If the user selects a specific peptide in the first tab, the values of the chosen categories for that peptide will be highlighted in red in the plot(s). The third tab 'Dynamic Scatter Plot' gives an overview of characteristics of all candidates in the dataset. Users can choose the variables to plot in x and y axis, as well as the variable which defines the size of the scatter plot. The variables can be transformed and limited in range, if desired. As the user hovers the cursor over any candidate, all information tied to the candidate will be displayed. With these features, users can quickly and interactively narrow down candidates satisfying criteria of interest. A curated subset of NeoFox scores that we believe are particularly useful and/or complementary to that provided by pVACtools are indicated in the pVACview NeoFox data exploration module.

**Custom module**

**Data import**
Users can also supply pVACview with any tsv file from any neoantigen prediction algorithm or pipeline. pVACview will read each column in the tsv as a feature and further tailors the view based on user's selected options in the following drop-down menus: (1) Group peptides by: this will group peptides together by a user-selected feature. For example, grouping by variant would consolidate all candidate peptides derived from a common variant. (2) Sort peptides by: this will order the candidate peptides by a user-selected feature. For example, a user might order peptides by binding score. (3) Features to display for each group of peptides: this menu is used to select which features in the dataset will be included in the detailed data section. By default, all features, with exception of the features chosen to group and sort peptide by, will be included. To demonstrate the custom input mode, we provide users with example results from three other neoantigen prediction pipelines: vaxrank [24], Neopredpipe [27], and antigen.garnish [28].

**Neoantigen visualization and exploration**
The custom module of pVACview offers three tabs for data visualization. The first tab 'Overview of Neoantigen Features' will feature groups of peptides. For each group, a single representative peptide will be shown. To see and compare the representative peptide with other peptides in the same group, users can click 'Investigate', and see all peptides in the second tab - 'Detailed Data' tab. In this second tab, the peptides in the group by default will be ordered by the feature users have selected to sort peptides by. The third tab 'Dynamic Scatter Plot' allows users to quickly and interactively narrow down candidates satisfying criteria of interest (as described in the NeoFox module section above).

As described above, pVACview provides a complex interactive interface to explore many neoantigen features and prioritize neoantigen candidates. A comprehensive analysis of the biological rationale and relative importance of individual features is beyond the scope of this report but several reviews and detailed guidelines have been published. In addition, we provide a list of suggested features and a brief description of their use in candidate prioritization as **Table 1**. More extensive discussions of many of these features are provided in instructional videos and a comprehensive vignette available in the online documentation.

## Results and Discussion

Multiple tools and workflows have been created for neoantigen characterization for studies of tumor T cell biology and the development of neoantigen-based therapies. Some existing tools that address individual factors for prediction of neoantigens do include visualization components (such as netMHC for binding predictions). Pipelines such as pVACtools and NeoFox combine multiple algorithms for detecting, characterizing and prioritizing neoantigens from various sources. However, these pipelines do not facilitate visual exploration of the results, instead producing a static neoantigen report that often attempts to provide only a single "best" neoantigen for each variant based on simple criteria such as binding affinity predictions. During this process, these reports over-simplify the outputs to make them tractable, and thus the results are fraught with assumptions about what "best" means. This reduces the ability to effectively prioritize neoantigen candidates. A more nuanced approach that allows consideration of many more contextually relevant features is more valuable. We therefore believe that an interactive neoantigen visualization tool, customized to this specific application is needed. Two existing tools, NeoPredViz [27] and LENS [29] do offer visualization of their results. However, pVACview remains distinct in its ability to present a diversity of variant, transcript and peptide data together in a simple, integrated view. As a dynamic interactive visualization interface, pVACview overcomes many limitations of tabular reports, allowing the user to consider neoantigens in the context of transcript expression, tumor clonality, multiple registers (peptides of the same length where the variant is at different positions), peptide lengths, alternative transcript isoforms, an ensemble of predictive binding algorithms, HLA specific anchor information and much more.

pVACview helps users to address many complexities of neoantigen interpretation and prioritization that are difficult to achieve with tabular reports. While there has been a rapid development of sequencing technologies, bulk tumor tissue samples undergoing sequencing are often subject to purity issues. Additionally, intratumoral heterogeneity presents a considerable challenge to cancer therapies, making it critical to distinguish variants from the founding clone from those that are subclonal. Neoantigens arise from tumor-specific genomic variations and each variant can have multiple transcripts encompassing the variant location. Thus, not only should transcript-specific expression level be taken into account, transcripts resulting from different splicing patterns may also have considerable impact on the exact neoantigen peptide sequence. Once the correct sequence surrounding a variant from an expressed transcript is identified, neoantigen candidates can be extracted by looking at different registers and different lengths of peptides containing the amino acid modification. Even for the simplest form of neoantigen sources, single nucleotide variants, when all registers, lengths (8-, 9-, 10-, 11-mer) and algorithms (13 for class I) in pVACtools are used, and assuming an individual with 6 distinct class I alleles, the results is 38 distinct short peptides and 2,964 peptide MHC predictions. Neoantigens arising from frameshift variants can produce an even larger number of neoantigen candidates that vary in sequence and variant position, with each peptide having a different set of prediction scores. Filtering by binding affinity thresholds or other criteria can reduce the complexity of this result to a degree but it often remains unwieldy. The detailed information provided by pVACview can effectively help users address this complexity.

Furthermore, pVACview's drill down approach to information display helps researchers integrate variant clonality, Class I and II binding predictions, competing binding prediction algorithms, binding metrics and mutation positional information for each candidate neoantigen peptide. Expressed neoantigens of different lengths and registers from a founding clone of the tumor can potentially bind (or not bind) to either Class I or Class II HLA alleles, either through endogenous or exogenous pathways respectively on either the tumor cell or an antigen presenting cell. In addition to having up to 6 different Class I HLA alleles, each patient may have up to 12 different Class II alleles (and dimer combinations of these). When evaluating neoantigen candidates in terms of binding predictions, researchers may take into account how well neoantigens are potentially binding to each patient-specific HLA allele and may also want to consider how many different HLA alleles it can bind robustly [30,31]. pVACview provides the ability to consider the diversity of peptides arising from each variant and how they relate to predicted binding by multiple alleles. Another approach to accounting for multiplicity of presentation is the Patient Harmonic-mean Best Rank (PHBR) score [32,33], conveniently provided by NeoFox and supported in pVACview. As described, for each neoantigen, there are numerous prediction algorithm results available. How the scores are distributed across different algorithms and whether the IC50 binding prediction or percentile rank value should be utilized are all important aspects that require careful evaluation. If a neoantigen is expressed and predicted to bind well to an HLA allele, researchers should further consider the anchor and mutation locations of the peptide-MHC pairing. A subset of peptide positions are primarily presented to the T-cell receptor for recognition, while others are primarily responsible for anchoring to the MHC (though neither role is exclusive). Whether the mutation lies in an anchor region and how well the WT peptide binds to the MHC creates different scenarios that can influence whether a neoantigen remains a good candidate [21,23].

Finally, in the setting of clinical trials, additional details influence the quality of neoantigens. Tumor samples are first collected from patients and put through a series of genomic analysis pipelines, including DNA and RNA sequencing, variant detection and expression estimation. Pipelines such as pVACtools then take these results and identify and characterize possible neoantigen candidates. Throughout this process, problems such as low tumor purity, contamination, and insufficient or excessive neoantigen candidates may arise. pVACview allows users to promptly adjust tiering of candidates based on tumor purity and expression levels. It also highlights the specific failing criteria, providing users the option to further explore criteria such as how to define anchor positions, with the flexibility of rescuing candidates for samples with insufficient candidates for downstream applications such as neoantigen vaccines. For cases where an excessive amount of neoantigen candidates exist, pVACview effectively prioritizes candidates (based on calculated tiering, allele expression and average mutant peptide binding affinity) while simultaneously allowing users to sort and annotate candidates based on features of their own choosing.

All the aspects described above are potentially critical in order to infer whether the presenting peptide sequence can successfully induce an immune response. pVACview was designed to present this complex information to researchers in an intuitive manner and aid in the prioritization and selection of neoantigen candidates for personalized cancer vaccines or other therapeutic and research applications. Basic and translational researchers can use pVACview to visualize neoantigen candidates along with detailed supporting information including that of the genomic variant, transcripts affected by the variant, and good-binding peptides predicted from the respective transcripts.

## Conclusions

Accurate neoantigen prediction is critical to cancer immunotherapy treatments and several tools have been built to account for individual aspects throughout this process. However, these tools lack methods for integration and visualization, making it challenging for researchers to efficiently explore the many molecular and algorithmic features relevant to neoantigens, such as variant, transcript and peptide level information. pVACview integrates multiple levels of information into a visualization tool, allowing users to analyze each candidate in detail for optimal decision making. This tool has been tested and used in clinical trials and research projects involving human, mouse, and canine model systems. We hope by using pVACview, researchers can analyze and prioritize neoantigen candidates with greater efficiency and accuracy. The application is available as part of the pVACtools pipeline and as an online web tool hosted on the Google Cloud Platform at [www.pvacview.org](www.pvacview.org).

## Availability and requirements

**Data availability**
The FASTQ and BAM files for HCC1395 and HCC1395BL demonstration data can be found at NCBI BioProject accession number: PRJNA201238 (SRA accession number: SRX285805, SRX285804, SRX278523, SRX278522, SRX278521, SRX278520, SRX278519, SRX278518, SRX278517).

**Software availability**
The pVACview codebase is hosted publicly on GitHub at https://github.com/griffithlab/pVACtools. User documentation is available at pvactools.org. This project is licensed under the BSD 3-Clause Clear License (https://github.com/griffithlab/pVACtools/blob/master/LICENSE). pVACtools has been packaged and uploaded to PyPI under the "pvactools" package name and can be installed on Linux or Mac systems by running the `pip install pvactools` command. Installation requires a Python 3.7 or higher environment which can be emulated by using Conda. Versioned Docker images including all dependencies are available on DockerHub (https://hub.docker.com/r/griffithlab/pvactools/). Releases are also made available on GitHub (https://github.com/griffithlab/pVACtools/releases).

## Acknowledgements

We thank the patients and their families for the donation of their samples for research purposes. Malachi Griffith was supported by the National Human Genome Research Institute (NHGRI) of the National Institutes of Health (NIH) under Award Number R00HG007940. Malachi Griffith and Obi Griffith were supported by the NIH National Cancer Institute (NCI) under Award Numbers U01CA209936, U01CA231844 and U24CA237719. Malachi Griffith was supported by the V Foundation for Cancer Research under Award Number V2018-007. William Gillanders was supported under the Centene Corporation contract (P19-00559) for the Washington University-Centene ARCH Personalized Medicine Initiative. Malachi Griffith, Obi Griffith, Huiming Xia, and William Gillanders were supported by the NCI under Award Number U01CA248235.

## Conflict of Interest

The authors declare that there is no conflict of interest.

## List of abbreviations

MT: Mutant
WT: Wild type
HLA: Human leukocyte antigen
MHC: Major histocompatibility complex
VCF: Variant allele frequency

## Figure 1

**Figure 1.** Personalized neoantigen prioritization and therapy development pipeline.
The process of developing personalized neoantigen therapy includes six main steps as depicted in this figure. The first step involves patient enrollment and collection of a tumor biopsy and matched normal sample. Next the samples undergo whole genome/exome and RNA sequencing, followed by variant calling to identify somatic variants unique to the cancer, as well as HLA typing. Information regarding the patients' variants and HLA type are fed to an ensemble of algorithms that predict neoantigen candidates. The candidates are then prioritized based on a multitude of criteria such as binding affinity, solubility, and variant expression. Finally, the selected candidates are sent to therapy manufacturers (e.g. peptide or nucleic acid cancer vaccines and adoptive cell therapies), subjected to safety testing, and ultimately delivered to the patient. pVACview is developed to aid the candidate prioritization and selection step.

## Figure 2

**Figure 2.** Overall process of neoantigen generation and presentation from tumor specific mutations.
Illustration demonstrating the key steps throughout the process of neoantigen generation, processing, binding, transportation and presentation. This overview highlights examples of criteria examined by pVACview that may be considered during neoantigen prioritization including: 1) founding clone versus subclonal tumor status of variants, 2) the impact of different transcripts structures on peptide sequence and their varying expression levels, 3) proteasomal processing generating peptides of varying length from different registers, 4) potential for peptide binding to individual patient MHC alleles and the predicted immunogenicity of peptide:MHC complexes, 5) variant and anchor positions of the neoantigen when presented to T-cells.

## Figure 3

**Figure 3**. Overview of example workflow for prioritizing neoantigens using pVACview. pVACview can be broken down into three main sections: upload, visualize/explore, and export. When exploring the neoantigen candidates, users are presented with three levels of information: variant, transcript and peptide. This example workflow guides the user through critical questions that may be considered when prioritizing neoantigen candidates. Each section is accompanied by the corresponding feature in the pVACview interface.

## Table 1

**Table 1.** Summary of pVACview features that facilitate neoantigen prioritization

| Level | Feature | Suggestion |
|---|---|---|
| Variant | RNA Expr : Gene expression value for the annotated gene containing the variant.<br><br>RNA VAF : Tumor RNA variant allele frequency (VAF) at this position.<br><br>Allele Expr : RNA Expr * RNA VAF<br><br>RNA Depth : Tumor RNA depth at this position.<br><br>DNA VAF : Tumor DNA variant allele frequency (VAF) at this position. | Prioritize variant(s) with high tumor variant allele fraction (VAF) and allelic expression. Interpretation of tumor VAF first requires estimation of tumor purity which may be guided by the VAF of known tumor drivers.<br><br>Prioritize variants in genes known to be associated with cancer type of the investigated sample, genes listed in Cancer Gene Census. |
| Transcript | TSL : Transcript support level of the best peptide | Prioritize transcript(s) where the annotated splicing pattern matches the splicing observed in patient RNAseq data (transcript with TSL >= 1, cross-check in IGV recommended). |
| Peptide | IC50 MT : Lowest or Median ic50 binding affinity of the best-binding mutant epitope across all prediction | Prioritize peptide(s) with favorable binding affinity to MHC (we recommend binding affinity of 500 nM or less, and percentile |

|  | algorithms used.<br><br>%ile MT : Lowest or Median binding affinity percentile rank of the best-binding mutant epitope across all prediction algorithms used (those that provide percentile output)<br><br>Elution score<br><br>Immunogenicity score | rank of 2 or less).<br><br>Prioritize peptides with favorable anchor scenarios. (see suggestion by Xia et al., 2023 [23])<br><br>Exclude peptides with a reference match in the proteome.<br><br>Prioritize peptides with favorable elution score (elution score ranges from 0 to 1, with 1 being the best elution score), with a percentile rank of 2 or less.<br><br>Prioritize peptides with favorable immunogenicity score (immunogenicity score from DeepImmuno and BigMHC_IM ranges from 0 to 1, with 1 being the best elution score). |
|---|---|---|

## Video Demonstration

We have created a series of demonstration videos that walk through the basic steps of using pVACview using neoantigen candidate files generated from the HCC1395 dataset as input. It shows the full process of launching the application, uploading datasets, exploring neoantigen information, adding comments and marking evaluations, and exporting the data for further usage. Additional videos describe modules customized to support NeoFox results or custom results from any neoantigen prediction pipeline:
https://www.youtube.com/channel/UCOJb2xPMhx2GSXiOfM4qmGw

**Figure 1.** Personalized neoantigen prioritization and therapy development pipeline.
The process of developing personalized neoantigen therapy includes six main steps as depicted in this figure. The first step involves patient enrollment and collection of a tumor biopsy and matched normal sample. Next the samples undergo whole genome/exome and RNA sequencing, followed by variant calling to identify somatic variants unique to the cancer, as well as HLA typing. Information regarding the patients' variants and HLA type are fed to an ensemble of algorithms that predict neoantigen candidates. The candidates are then prioritized based on a multitude of criteria such as binding affinity, solubility, and variant expression. Finally, the selected candidates are sent to therapy manufacturers (e.g. peptide or nucleic acid cancer vaccines and adoptive cell therapies), subjected to safety testing, and ultimately delivered to the patient. pVACview is developed to aid the candidate prioritization and selection step.

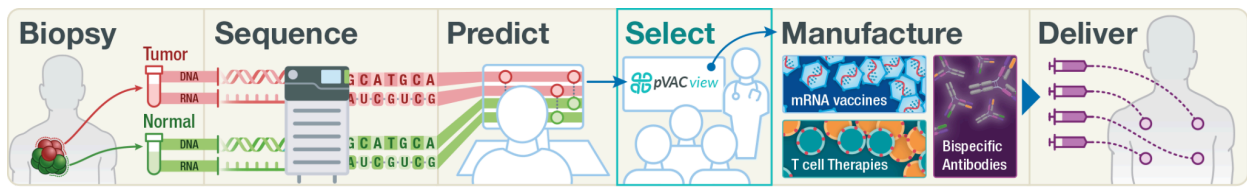

**Figure 2.** Overall process of neoantigen generation and presentation from tumor specific mutations.
Illustration demonstrating the key steps throughout the process of neoantigen generation, processing, binding, transportation and presentation. This overview highlights examples of criteria examined by pVACview that may be considered during neoantigen prioritization including: 1) founding clone versus subclonal tumor status of variants, 2) the impact of different transcripts structures on peptide sequence and their varying expression levels, 3) proteasomal processing generating peptides of varying length from different registers, 4) potential for peptide binding to individual patient MHC alleles and the predicted immunogenicity of peptide:MHC complexes, 5) variant and anchor positions of the neoantigen when presented to T-cells.

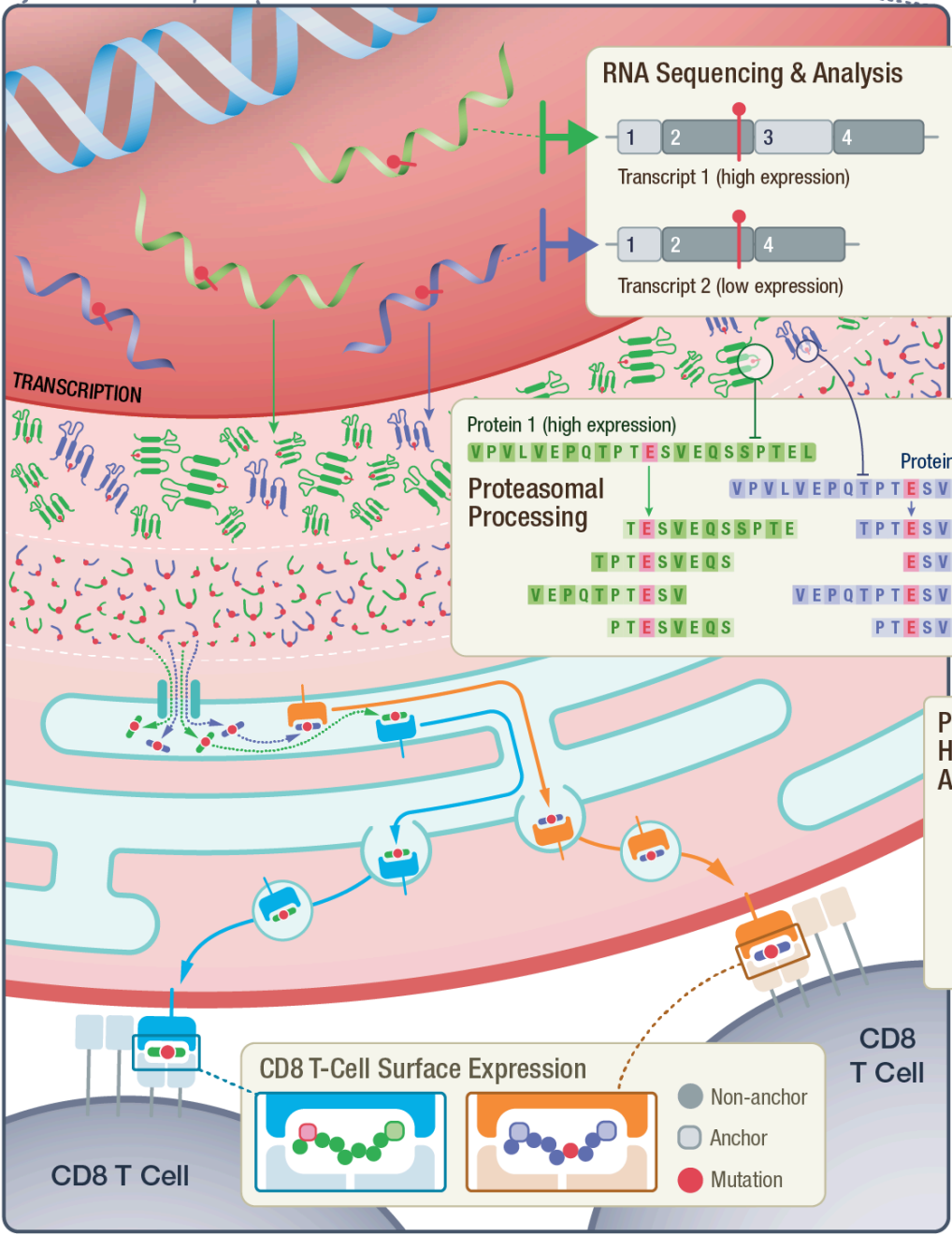

**Figure 3**. Overview of example workflow for prioritizing neoantigens using pVACview. pVACview can be broken down into three main sections: upload, visualize/explore, and export. When exploring the neoantigen candidates, users are presented with three levels of information: variant, transcript and peptide. This example workflow guides the user through critical questions that may be considered when prioritizing neoantigen candidates. Each section is accompanied by the corresponding feature in the pVACview interface.

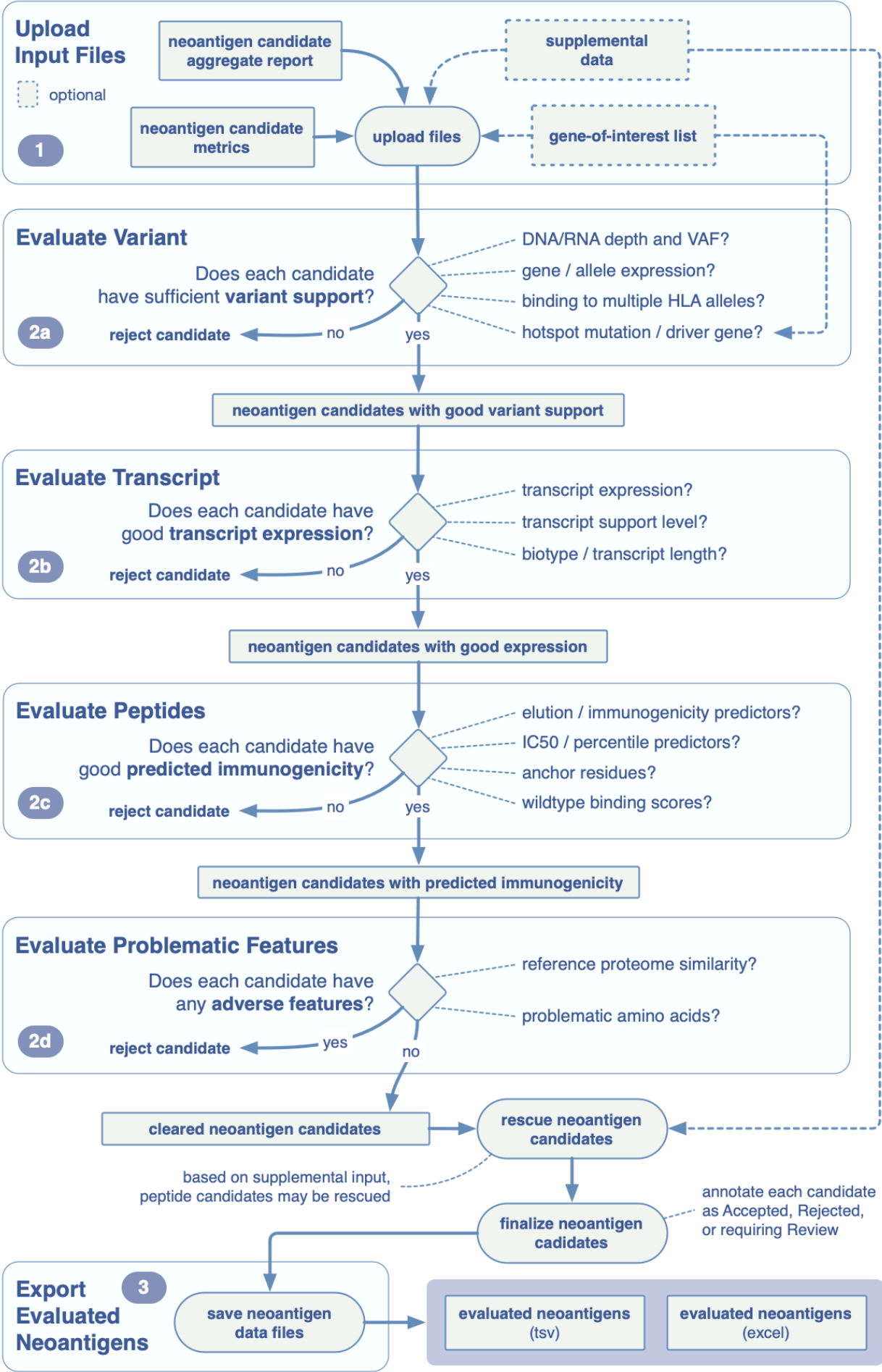